\documentclass[10pt]{article}
\usepackage{graphicx}

\newcommand{\Br}{{\bf r}}
\newcommand{\Brho}{\mbox{\boldmath $\rho$}}
\renewcommand{\Im}{{\rm Im}}

\newcommand{\nm}{\mbox{ nm}}
\newcommand{\mm}{\mbox{ mm}}
\newcommand{\m}{\mbox{ m}}

\begin{document}

%\preprint{}

\title{Correlation-induced orbital angular momentum changes}

\author{Yongtao Zhang \\ College of Physics and Information Engineering \\ Minnan Normal University, Zhangzhou 363000, China \\ Department of Physics and Optical Science \\ University of North Carolina at Charlotte, Charlotte, NC 28277, USA \\ Olga Korotkova \\ Department of Physics \\ University of Miami, Coral Gables, Florida 33146, USA \\ Yangjian Cai \\ Shandong Provincial Engineering and Technical Center of Light Manipulations \\ \& Shandong Provincial Key Laboratory of Optics and Photonic Device \\ School of Physics and Electronics \\ Shandong Normal University, Jinan 250358, China \\ School of Physical Science and Technology \\ Soochow University, Suzhou 215006, China \\ Greg Gbur \\ Department of Physics and Optical Science \\ University of North Carolina at Charlotte, Charlotte, NC 28277, USA}

\maketitle


\date{}

\begin{abstract}
We demonstrate that the orbital angular momentum flux density of a paraxial light beam can change on propagation in free space. These changes are entirely due to the spatial coherence state of the source, and the effect is analogous to correlation-induced changes in the intensity, spectrum and polarization of a beam. The use of the source coherence state to control the width, shape, and transverse shifts of the OAM flux density is demonstrated with numerical examples.
\end{abstract}

%\keywords{}


Over the development of the theory of optical coherence, it has been increasingly appreciated that the state of spatial coherence can influence a number of fundamental properties of a light beam on propagation in free space, without any interactions with a medium.  The earliest example of this was introduced in work by Collett and Wolf in 1978, in which they demonstrated that the diffraction rate of a partially coherent beam can be controlled by the source coherence state \cite{ecew:ol:1978}.  In the late 1980s, Wolf and colleagues caused a scientific furor when they showed that the spectrum of light can be changed on propagation due to source coherence \cite{ew:prl:1986,zdew:josaa:1988}, and can even under special circumstances mimic cosmological redshifts \cite{ew:n:1987}.  This revelation led to searches for other correlation-induced changes in light beams. In 1994, James demonstrated that source coherence may be responsible for changes in the degree of polarization \cite{dfvj:josaa:1994}, and in 2005 Korotkova and Wolf showed that the state of polarization may be similarly affected \cite{okew:oc:2005}.  In this letter we introduce another effect related to the source spatial coherence: correlation-induced changes in the orbital angular momentum of light.

In recent years, there has been significant interest in the development of singular optics, as a relatively unexplored area of classical electromagnetism, for use in innovative technologies \cite{soskin2001singular,dennis2009singular,gbur2016singular}.  This research area largely focuses on optical vortices (singularities of wavefield phase), such as those which appear in Laguerre-Gauss beams of nonzero azimuthal order. These vortex beams are now known to possess orbital angular momentum (OAM), which is connected to the circulating phase of the vortex \cite{allen1992orbital,lasmbmjp:oam:2003}.  The OAM of light has already been applied to a number of applications, including optical trapping and rotation \cite{simpson1997mechanical}, the design of light-driven machines \cite{ladavac2004microoptomechanical}, and free-space optical communication \cite{gibson2004free,jwyyyimfnayyhhyryysdmtaew:np:2012}.

The relationship between partial coherence and OAM has only recently been explored in significant detail.  Early work demonstrated that optical vortices turn into vortices of the two-point correlation functions as the spatial coherence is decreased \cite{gbur2003coherence,gbur2004hidden,palacios2004spatial}, suggesting that such structures are relatively robust.  The OAM of partially coherent vortex beams, in particular the distribution of OAM flux in a beam's cross-section, was only investigated in 2012 \cite{kim2012angular}, and it was shown that it generally manifests a Rankine vortex structure \cite{gasjriha:prl:2007}, with a core rotating like a rigid body and outskirts rotating like a fluid. In 2018, Gbur showed that there exist three fundamental classes of partially coherent vortex beams, characterized by their OAM distribution: rigid rotators, fluid rotators, and Rankine rotators \cite{gg:pspie:2018}.  

In this letter, we show that one can use source correlations to dramatically change the OAM flux density of a beam on propagation in free space.  These correlation-induced OAM changes can be quite dramatic and can be controlled by an appropriate choice of source correlations.  Since the earlier work on correlation-induced spectral changes and polarization was done, coherence theory itself has evolved considerably, with new classes of correlation functions providing many novel effects and tunability. For example, it has been shown that structuring of the amplitude \cite{sahin2012light} or phase \cite{okxc:ol:2018} of the source coherence state can result in controllable beam profiles, tilts, radial acceleration or asymmetric splitting of a propagating beam.  Here we derive the basic theory showing how source coherence results in unusual distributions of OAM flux density on propagation, and illustrate such changes with a number of source coherence models.

A scalar partially coherent beam in a plane perpendicular to the direction of propagation can be characterized by the cross-spectral density (CSD) function \cite[Chapter 4]{wolf2007introduction},
\begin{equation}
W(\Br_1,\Br_2) = \langle U^\ast(\Br_1)U(\Br_2)\rangle_\omega,
\end{equation}
where $U(\Br)$ is the field at frequency $\omega$, $\Br_1$ and $\Br_2$ are the transverse position vectors, and the angular brackets $\langle \cdots \rangle_\omega$ represent averaging over a space-frequency ensemble \cite[Chapter 4]{wolf2007introduction}.

The average OAM flux density $L_d(\Br)$ of such a beam can be shown to be related to the CSD by the expression \cite{kim2012angular},
\begin{equation}
L_d(\Br) = -\frac{\epsilon_0}{k}\Im \left\{(y_1 \partial_{x_2}-x_1\partial_{y_2})W(\Br_1,\Br_2)\right\}_{\Br_1=\Br_2=\Br},
\end{equation}
where $k=2\pi/\lambda$ is the wavenumber of the light, with $\lambda$ being the wavelength, and $\epsilon_0$ is the free-space permittivity.  Here $\partial_{\alpha_2}$ represents the partial derivative with respect to $\alpha_2$, with $\alpha=x,y$.

The magnitude of the OAM flux density at a particular point in space arises from two distinct effects: the average OAM per photon and the density of photons in space, i.e. the intensity.  We isolate the OAM of individual photons by considering the average normalized OAM flux density, defined as
\begin{equation}
l_d(\Br) = \frac{\hbar \omega L_d(\Br)}{S(\Br)}, 
\end{equation}
where 
\begin{equation}
S(\Br) = \frac{k}{\mu_0\omega}W(\Br,\Br)
\end{equation}
is the $z$-component of the Poynting vector. The dimensions of $l_d(\Br)$ are angular momentum per photon; the quantity therefore represents the average OAM of a photon measured at the particular position $\Br$.  For any Laguerre-Gauss mode of azimuthal order $l$, we have $l_d(\Br)=l\hbar$, a constant. 

We may also introduce the total OAM per photon $t_d$ by the related formula,
\begin{equation}
t_d = \frac{\hbar \omega \int L_d(\Br)d^2r}{\int S(\Br)d^2r}.
\end{equation}
We now demonstrate that while $t_d$ remains constant on propagation, as total OAM is conserved, the quantity $l_d(\Br)$ may exhibit dramatic changes on propagation, which for our examples are entirely due to the source coherence.

The CSD of a source in the plane $z=0$ propagates in free space to a distance $z$ via Fresnel diffraction,
\begin{eqnarray}
&&W(\Brho_1,\Brho_2,z) = \frac{1}{(\lambda z)^2}\int\int W(\Br_1,\Br_2) \nonumber \\
&\times&\exp\left\{-\frac{ik}{2z}\left[(\Br_1-\Brho_1)^2-(\Br_2-\Brho_2)^2\right]\right\}d^2r_1d^2r_2,
\label{W:prop}
\end{eqnarray}
where $\Brho_1$, $\Brho_2$ are the transverse position vectors at distance $z$.

We begin by considering isotropic Schell-model sources with a definite topological charge, of the form
\begin{equation}
W(\Br_1,\Br_2) = U_l^\ast(\Br_1)U_l(\Br_2)\mu_0(|\Br_2-\Br_1|), \label{firstexamples}
\end{equation}
where $\mu_0(|\Br_2-\Br_1|)$ is the spectral degree of coherence of the field with charge $l$, and the function $U_l(\Br)$ represents a vortex beam of radial order $0$ and azimuthal order $l$, expressed in polar coordinates $(r,\phi)$ as
\begin{equation}
U_l(\Br) = C_l r^{|l|} \exp[il\phi] \exp\left[-\frac{r^2}{w^2}\right].
\end{equation}
In this expression, we have
\begin{equation}
C_l \equiv \sqrt{\frac{2}{\pi w^2|l|!}}\left(\frac{\sqrt{2}}{w}\right)^{|l|},
\end{equation}
with $\Br = (r,\phi)$, and $w$ is the initial beam width.

Because the spectral degree of coherence is homogeneous and isotropic, it has no effect on the total OAM $t_d = l\hbar$ or the source distribution of OAM, i.e. $l_d(\Br)=l \hbar$.  If the field were fully coherent, $l_d(\Br)$ would also remain constant on propagation -- any change observed in a partially coherent beam of the form of Eq.~(\ref{firstexamples}) is therefore entirely due to the effects of partial coherence.

To show the variety of changes possible, we consider a multi-Gaussian Schell-model vortex (MGSMV) source \cite{sahin2012light} with a source spectral degree of coherence
\begin{equation}
\mu_0(|\Br_2-\Br_1|) = \frac{1}{C_0}\sum_{m=1}^\infty \frac{(-1)^{m-1}}{m}\left(\begin{array}{c} M \\ m \end{array}\right) \exp\left[-\frac{|\Br_1-\Br_2|^2}{\delta^2}\right],
\end{equation}
where
\begin{equation}
C_0 \equiv \sum_{m=1}^\infty\frac{(-1)^{m-1}}{m}\left(\begin{array}{c} M \\ m \end{array}\right)
\end{equation}
is a normalization factor, $\left(\begin{array}{c} M \\ m\end{array}\right)$ is a generalized binomial coefficient which can take integer or fractional values of $M$ \cite{fwok:oe:2017}, and $\delta_m = \sqrt{m} \delta$ are the correlation widths of the constituent Gaussian functions. For $M>1$, this degree of coherence produces a flat-topped beam profile; for $M<1$, it produces a cusped beam profile. For $M=1$, this reduces to the traditional Gaussian Schell-model case, i.e.
\begin{equation}
\mu_0(|\Br_2-\Br_1|) = \exp\left[-\frac{|\Br_1-\Br_2|^2}{\delta^2}\right],
\end{equation}
where $\delta$ is the correlation width.

We consider the case $l = 1$; on substitution from Eq.~(\ref{firstexamples}) into Eq.~(\ref{W:prop}), we can analytically evaluate the CSD at any propagation distance; expressions for the CSD, the OAM flux density, and the Poynting vector are given in supplementary material.

Figure \ref{MGSMfig} shows the evolution of the normalized OAM flux density $l_d$ of such MGSMV beams on propagation, for different orders $M$ and different correlation widths $\delta$.  We take $\lambda = 632.8\nm$ and $w=1\mm$ for the remainder of the letter. Figures \ref{MGSMfig}(a) and (b) illustrate the effects of correlation width $\delta$, with $M=1$, the case of a Gaussian Schell-model beam.  At small propagation distances, the beams act largely like fluid rotators -- with near-constant $l_d(\Br)$ and a very small rigid body region in the core -- and the rigid body region expands as it propagates. A beam with lower spatial coherence (smaller $\delta$) will exhibit this rigid body behavior at a shorter propagation distance. These figures demonstrate how the correlation width dramatically affects the distribution of OAM on propagation.

\begin{figure}
\centering
\includegraphics[width=5in]{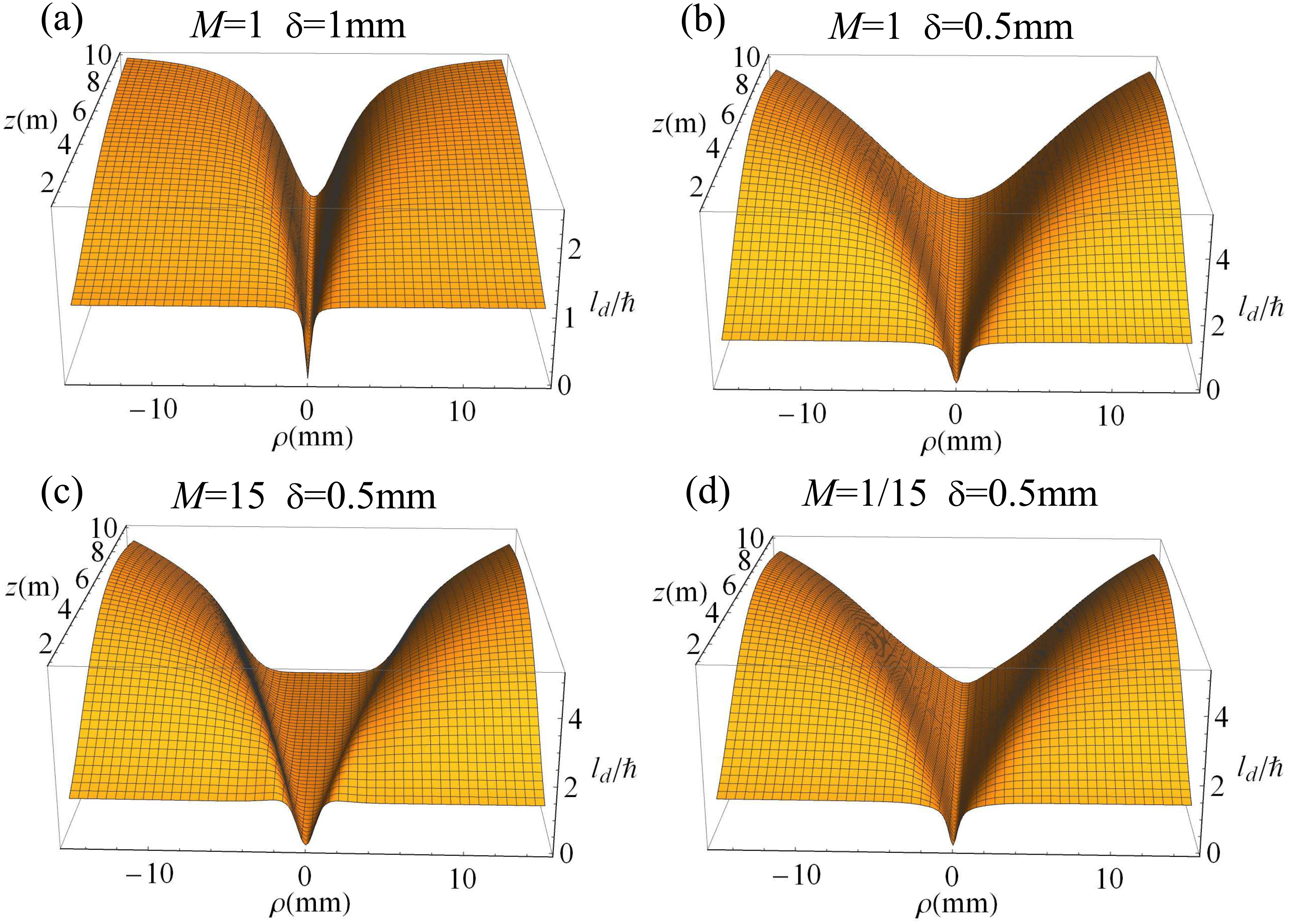}
\caption{The normalized OAM flux density of multi-Gaussian Schell-model beams on propagation, for different orders $M$ and different correlation widths $\delta$.}
\label{MGSMfig}
\end{figure}

Figures \ref{MGSMfig}(c) and (d) show examples of the normalized OAM flux density for $M>1$ and $M<1$; it can be seen that this flux density takes on flat-topped or cusped profiles, respectively.  The former case represents an OAM ``dead zone'' in the center of the beam, in which there is no circulation.  

\begin{figure}
\centering
\includegraphics[width=4in]{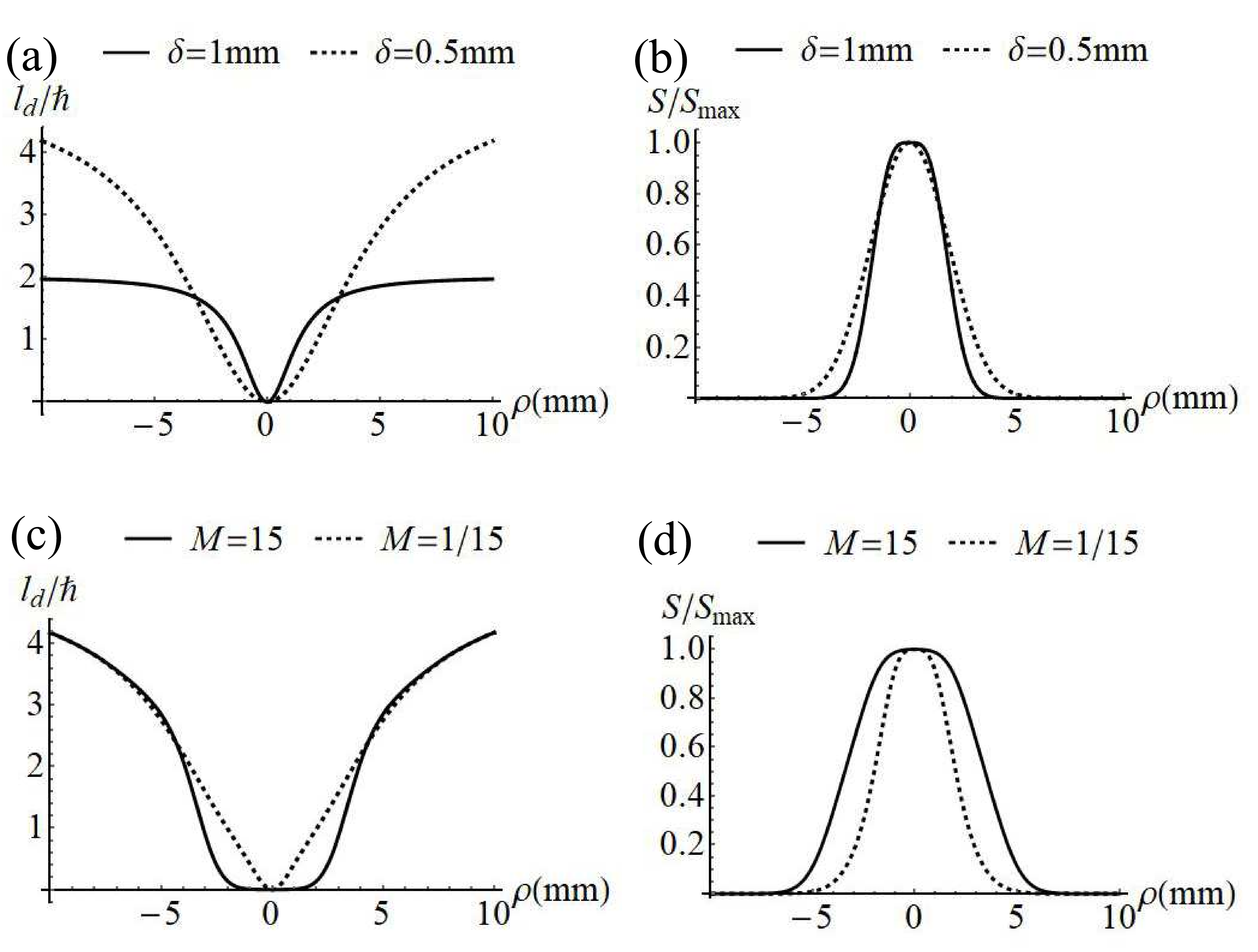}
\caption{The normalized OAM flux density and spectral density of MGSMV beams with $M=1$ and $z=5\m$ ((a) and (b)), and $\delta=0.5\mm$ and $z=5\m$ ((c) and (d)).}
\label{spectraldens}
\end{figure}

These latter examples, curiously, mimic the spectral density profiles of their respective beams: a beam with a flat-topped spectral density, for instance, results in a flat-bottomed ``dead zone'' in the OAM.  The OAM flux densities and corresponding spectral densities $S(\Br) = W(\Br,\Br)$ are illustrated for MGSM beams at a fixed propagation distance in Fig.~\ref{spectraldens}.  It is to be noted that the most dramatic changes in the OAM flux density occur in regions where the spectral density is still high: these OAM changes are not confined to regions with few or no photons.

\begin{figure}
\centering
\includegraphics[width=4in]{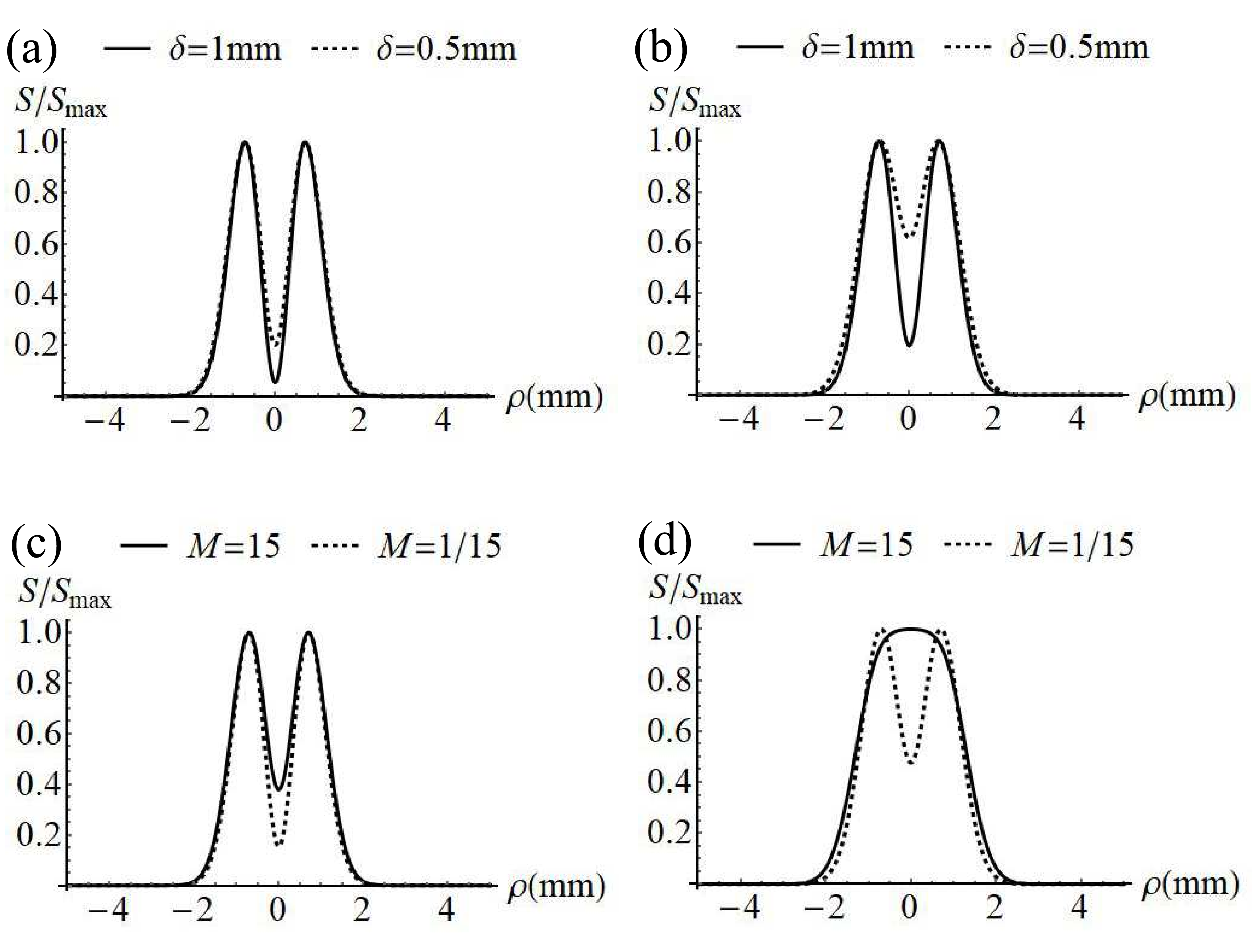}
\caption{The normalized spectral density of MGSMV beams at $z=0.5\m$ ((a) and (c)) and $z=1\m$ ((b) and (d)), with $M=1$ ((a) and (b)) and $\delta=0.5\mm$  ((c) and (d)).}
\label{spectraldens2}
\end{figure}

The propagation distances in Fig.~\ref{spectraldens} were chosen to highlight the most dramatic OAM flux changes; however, this does not clearly show the evolution of the spectral density for the different cases. In Fig.~\ref{spectraldens2} the spectral densities of all cases are shown at $z=0.5\m$ and $z=1.0\m$. It can be seen that all beams maintain a low intensity core at short propagation distances, which becomes less prominent and eventually vanishes at large propagation distances. This demonstrates that the OAM flux density changes are not simply an artifact of the distribution of spectral density.

\begin{figure}
\centering
\includegraphics[width=4in]{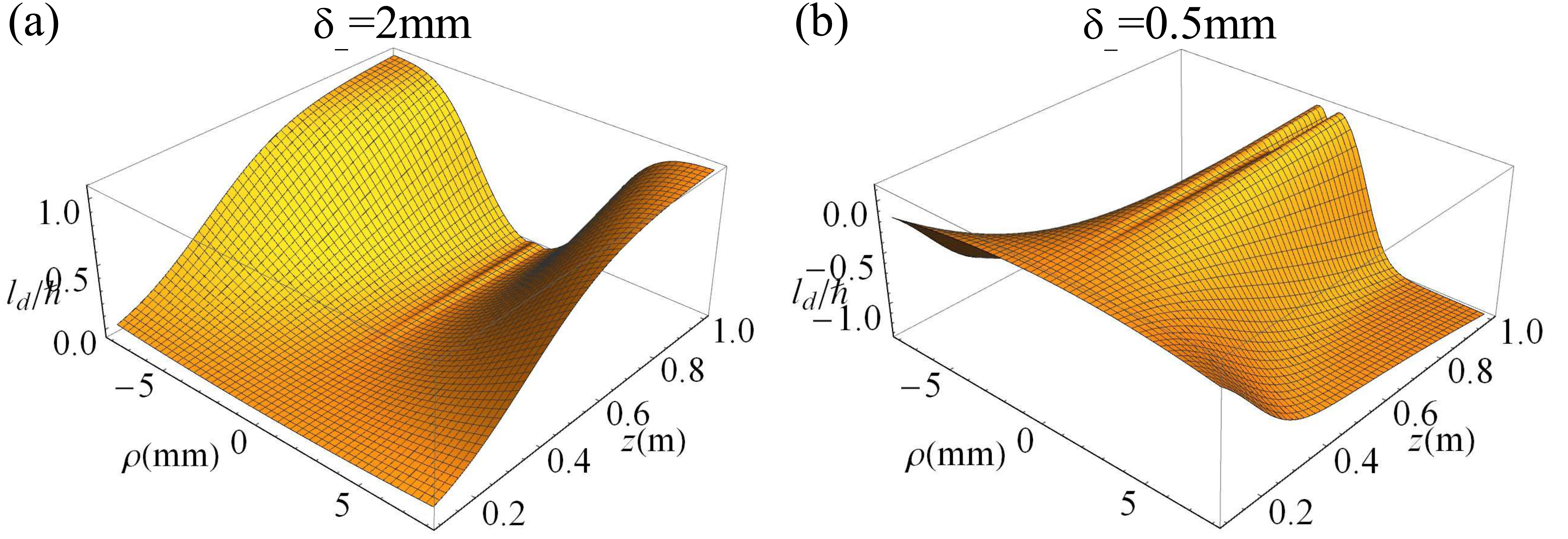}
\caption{The normalized OAM flux density of an incoherent superposition of GSMV beams with $|l|=1$ and $\delta_+=1\mm$.}
\label{OAMincoh}
\end{figure}

\begin{figure}
\centering
\includegraphics[width=4in]{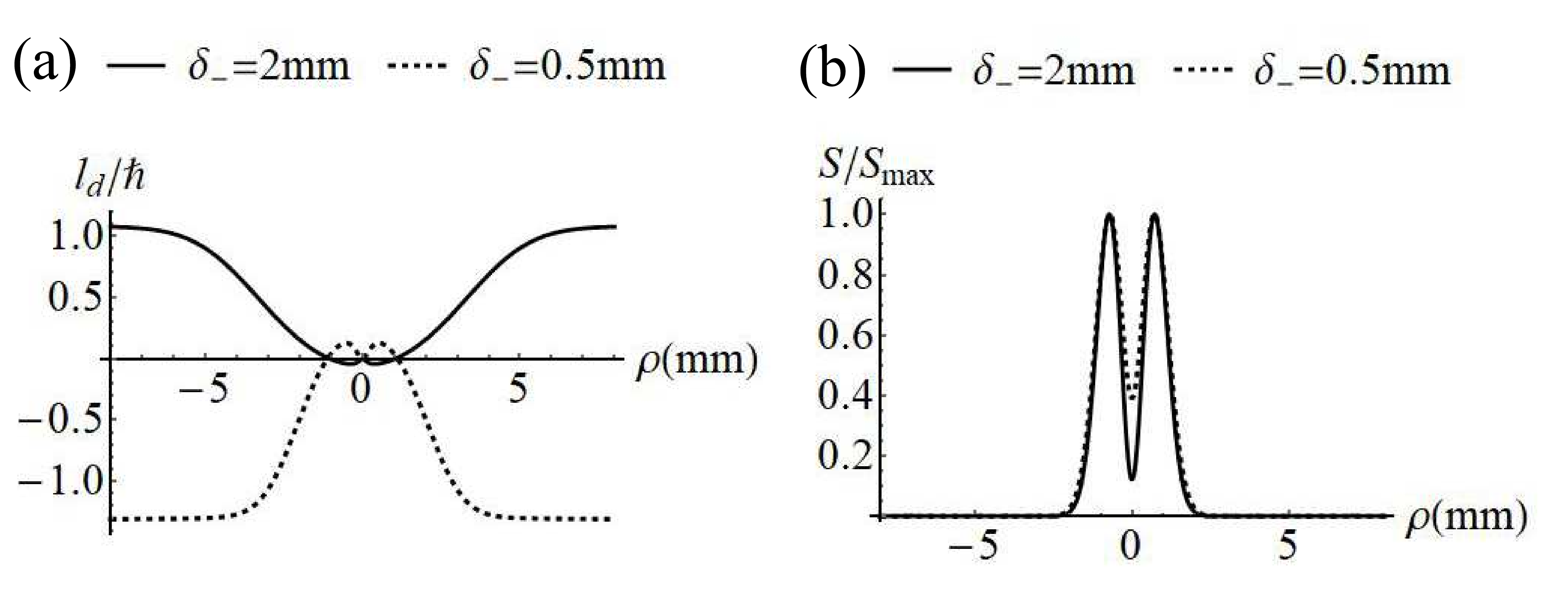}
\caption{The (a) normalized OAM flux density and (b) spectral density of an incoherent superposition of GSMV beams with $|l|=1$,  $\delta_{+} = 1\mm$ and $z=1\m$.}
\label{OAMdelta}
\end{figure}

Even more extreme changes in the distribution of OAM can be achieved with partial coherence, even the creation of counter-rotating regions.  To illustrate this, we consider a source which is an incoherent superposition of GSMV beams with equal and opposite topological charges but different correlation widths $\delta_n$, i.e.
\begin{equation}
W(\Br_1,\Br_2) = \sum_{n=+l, -l} U^\ast_{n}(\Br_1)U_{n}(\Br_2)\exp\left[-\frac{|\Br_2-\Br_1|^2}{\delta_n^2}\right].\label{twobeam}
\end{equation}
Because the intensity profiles of the two contributions are the same, $t_d = 0$ and the normalized OAM flux density $l_d(\Br) = 0$ in the source plane. But if the correlation widths of the two components are different, they will propagate differently, resulting in a non-constant normalized OAM flux density away from the source. The continuous evolution of $l_d(\Br)$ as a function of $z$ is shown in Fig.~\ref{OAMincoh}, for two different values of $\delta_-$, with $|l|=1$, $\delta_{+} = 1\mm$.  For both values of $\delta_-$, one can see a ``dent'' in the curve near $\rho=0$.  To highlight this, we plot the OAM flux density for several values of $\delta_-$ in Fig.~\ref{OAMdelta} at $z=1\m$.  We can see that there exist counter-rotating (positive and negative) regions of OAM flux density in the beam when $\delta_+$ and $\delta_-$ are significantly different from each other, and that the radial locations of these regions can be ``flipped'' by changing the value of $\delta_-$. The spectral densities for the examples are shown in Fig.~\ref{OAMdelta}(b).

It is natural to ask how one might directly measure these OAM changes experimentally. The most straightforward possibility would be to measure the motion of microscopic particles in the path of the beam, deducing the strength and handedness of the rotation by the direction the particle moves. Small particles can be put into orbital motion by OAM beams; this was impressively demonstrated some years ago when microspheres were used in a sequence of counter-rotating vortex beams to create a microoptomechanical fluid pump \cite{ladavac2004microoptomechanical}. A beam such as that described by Eq.~(\ref{twobeam}), which produces counter-rotating regions from a beam that starts with no net OAM, would provide a clear qualitative indication of correlation-induced OAM changes. 

We have therefore demonstrated the existence of correlation-induced OAM changes, which join spectral changes and polarization changes in a family of coherence-influenced propagation phenomena.  This work, which highlights a previously unexplored physical phenomenon, also provides an additional degree of freedom for the control and trapping of microscopic particles.

The authors would like to acknowledge the National Key Research and Development Project of China (Grant No. 2019YFA0705000), National Natural Science Foundation of China (NSFC) (Grant Nos. 11974218, 91750201, 11525418, 11474143), Innovation Group of Jinan (2018GXRC010).      
%\end{acknowledgments}

\bibliographystyle{abbrv}
\bibliography{correlation}

\end{document}